\documentclass[preprint,superscriptaddress]{revtex4-1}
\usepackage{graphicx}
\usepackage{braket}
\usepackage{amssymb,amsmath}
\usepackage[T1]{fontenc}
\usepackage[english]{babel}
\usepackage{textgreek}

\begin{document}

\title{High-Gain Harmonic Generation with temporally overlapping seed pulses and application to ultrafast spectroscopy}

\author{Andreas Wituschek}
\email{andreas.wituschek@physik.uni-freiburg.de}
\affiliation{Institute of Physics, University of Freiburg, Hermann-Herder-Str. 3, 79104 Freiburg, Germany}
\author{Lukas Bruder}
\email{lukas.bruder@physik.uni-freiburg.de}
\affiliation{Institute of Physics, University of Freiburg, Hermann-Herder-Str. 3, 79104 Freiburg, Germany}
\author{Enrico Allaria}
\affiliation{Elettra-Sincrotrone Trieste S.C.p.A., 34149 Basovizza (Trieste), Italy.}
\author{Ulrich Bangert}
\author{Marcel Binz}
\affiliation{Institute of Physics, University of Freiburg, Hermann-Herder-Str. 3, 79104 Freiburg, Germany}
\author{Carlo Callegari}
\author{Paolo Cinquegrana}
\author{Miltcho Danailov}
\author{Alexander Demidovich}
\author{Michele Di Fraia}
\affiliation{Elettra-Sincrotrone Trieste S.C.p.A., 34149 Basovizza (Trieste), Italy.}
\author{Raimund Feifel}
\affiliation{Department of
Physics, University of Gothenburg, Origov\"agen 6 B, SE-412 96 Gothenburg, Sweden}
\author{Tim Laarmann}
\affiliation{Deutsches Elektronen-Synchrotron DESY, Notkestr. 85, 22607
Hamburg, Germany}
\affiliation{The Hamburg Centre for Ultrafast Imaging CUI, Luruper Chaussee 149, 22761 Hamburg, Germany}
\author{Rupert Michiels}
\affiliation{Institute of Physics, University of Freiburg, Hermann-Herder-Str. 3, 79104 Freiburg, Germany}
\author{Marcel Mudrich}
\affiliation{Department of Physics and
Astronomy, Aarhus University, Ny Munkegade 120, DK-8000 Aarhus, Denmark}
\author{Ivaylo Nikolov}
\affiliation{Elettra-Sincrotrone Trieste S.C.p.A., 34149 Basovizza (Trieste), Italy.}
\author{Paolo Piseri}
\affiliation{Dipartimento di Fisica `Aldo Pontremoli', Universit\`{a} degli Studi di Milano, via Celoria 16, 20133, Italy}
\author{Oksana Plekan}
\author{Kevin Charles Prince}
\affiliation{Elettra-Sincrotrone Trieste S.C.p.A., 34149 Basovizza (Trieste), Italy.}
\author{Andreas Przystawik}
\affiliation{Deutsches Elektronen-Synchrotron DESY, Notkestr. 85, 22607
Hamburg, Germany}
\author{Primo\v{z} Rebernic Ribi\v{c}}
\affiliation{Elettra-Sincrotrone Trieste S.C.p.A., 34149 Basovizza (Trieste), Italy.}
\affiliation{Laboratory of Quantum Optics, University of Nova Gorica, 5001 Nova Gorica, Slovenia}
\author{Paolo Sigalotti}
\affiliation{Elettra-Sincrotrone Trieste S.C.p.A., 34149 Basovizza (Trieste), Italy.}
\author{Stefano Stranges}
\affiliation{Department of Chemistry and Drug Technologies,
University of Rome `Sapienza', P.le A. Moro 5, 00185 Roma, Italy}
\affiliation{IOM-CNR Tasc Laboratory, Strada Statale 14 km 163.5 34149 Basovizza (Trieste), Italy}
\author{Daniel Uhl}
\affiliation{Institute of Physics, University of Freiburg, Hermann-Herder-Str. 3, 79104 Freiburg, Germany}
\author{Luca Giannessi}
\affiliation{Elettra-Sincrotrone Trieste S.C.p.A., 34149 Basovizza (Trieste), Italy.}
\affiliation{Istituto Nazionale di Fisica Nucleare - Laboratori Nazionali di Frascati, Via E. Fermi 40, 00044 Frascati, Roma}
\author{Frank Stienkemeier}
\affiliation{Institute of Physics, University of Freiburg, Hermann-Herder-Str. 3, 79104 Freiburg, Germany}

\date{July 24, 2020}

\begin{abstract}
Collinear double-pulse seeding of the High-Gain Harmonic Generation (HGHG) process in a free-electron laser (FEL) is a promising approach to facilitate various coherent nonlinear spectroscopy schemes in the extreme ultraviolet (XUV) spectral range. However, in collinear arrangements using a single nonlinear medium, temporally overlapping seed pulses may introduce nonlinear mixing signals that compromise the experiment at short time delays. Here, we investigate these effects in detail by extending the analysis described in a recent publication (Wituschek et al., Nat. Commun., \textbf{11}, 883, 2020). High-order fringe-resolved autocorrelation and wave packet interferometry experiments at photon energies > 23 eV are performed, accompanied by numerical simulations. 
It turns out that both the autocorrelation and the wave-packet interferometry data are very sensitive to saturation effects and can thus be used to characterize saturation in the HGHG process. Our results further imply that time-resolved spectroscopy experiments are feasible even for time delays smaller than the seed pulse duration.
\end{abstract}

\maketitle


\section{Introduction}
The extension of coherent nonlinear spectroscopy techniques to the extreme ultraviolet (XUV) and x-ray spectral regimes would allow the study of photoinduced dynamics in real-time with unprecedented temporal resolution and site/chemical selectivity\,\cite{ramasesha_real-time_2016,mukamel_multidimensional_2013,kraus_ultrafast_2018}.
However, for this development, the generation and control of phase-locked XUV/x-ray pulse sequences and selective background-free detection of weak nonlinear signals is essential\,\cite{mukamel_principles_1999}.
Background-free detection was demonstrated in four wave mixing (FWM) studies involving XUV excitation\,\cite{bencivenga_four-wave_2015,marroux_multidimensional_2018,ding_time-resolved_2016}.
However, non-collinear beam geometries are limited by mechanical instabilities which impedes precise interferometric measurements of electronic coherences in the XUV spectral range. 
The use of diffractive optics can avoid instabilities, but restricts the possibility to control pulse delays\,\cite{svetina_towards_2019}. 
In addition, due to the detection of photons, the overall sensitivity of FWM schemes is limited by stray light. 

%
On the other hand, phase-locking in XUV pulse trains was achieved in several experiments\,\cite{prince_coherent_2016,usenko_attosecond_2017,okino_direct_2015, maroju_attosecond_2020}. 
In this way, Ramsey-type spectroscopy was demonstrated\,\cite{eramo_method_2011} and one-femtosecond electron dynamics was followed in real-time\,\cite{tzallas_extreme-ultraviolet_2011}.
However, these experimental schemes mostly lack the capability for selective background-free detection of nonlinear signals, e.g. the third order polarization.


We have recently implemented a phase-modulation technique for XUV pulses, which, for the first time, simultaneously achieved phase-locking and strong background suppression due to lock-in detection\,\cite{wituschek_tracking_2020}.
This allowed us to follow the coherent evolution of XUV wave packets with high time resolution and sensitivity.
To this end, we performed twin-seeding\,\cite{gauthier_generation_2016} of the High-Gain Harmonic Generation (HGHG) process with phase-modulated ultraviolet (UV) seed pulse pairs at the FERMI free-electron laser (FEL).
The phase-modulation technique was originally developed for electronic wave-packet interferometry (WPI) studies with near-infrared lasers\,\cite{tekavec_wave_2006}.
In this method, information about the coherent evolution of an electronic wave packet is gained by monitoring incoherent `action' signals\,\cite{tekavec_wave_2006,bruder_phase-modulated_2015,nardin_multidimensional_2013}.
In addition, quasi phase-cycling and phase-sensitive lock-in detection of the action signals is introduced, which permits rotating-frame detection and pathway selection while having all the benefits from a collinear beam geometry, e.g. easy implementation and maximum interference contrast. 
The quasi phase-cycling enables flexible signal selection protocols, for instance to design highly sensitive probes for interparticle interactions\,\cite{bruder_delocalized_2019} or to perform coherent multidimensional spectroscopy in the gas phase \,\cite{tekavec_fluorescence-detected_2007,bruder_coherent_2018}.

The implementation of XUV phase modulation via double-pulse seeding of HGHG has the advantage of avoiding the technical challenge of direct XUV pulse manipulation\,\cite{lazzarino_shaping_2019}. 
Instead, timing and phase properties of the seed pulses can be controlled to high precision acting on the UV pulses on the optical table\,\cite{wituschek_stable_2019,gauthier_generation_2016}. 
However, when the two seed pulses begin to overlap temporally, interference leads to nonlinear mixing in the HGHG process, which has an impact on the measured WPI signals.

In this work we investigate in detail the signal contributions during temporal overlap of twin-pulse seeding HGHG at the FERMI FEL.
We record fringe-resolved high-order interferometric autocorelation traces and investigate experimentally and theoretically the competition of nonlinear mixing signals with the linear response of excited He atoms at 23.7\,eV. 
We find that the phase modulation method is capable of extracting the system's response even for temporally overlapping seed pulses, however saturation occurring during the HGHG process leads to strong depletion of the WPI signal.

\begin{figure}
\centering\includegraphics[width=0.9\linewidth]{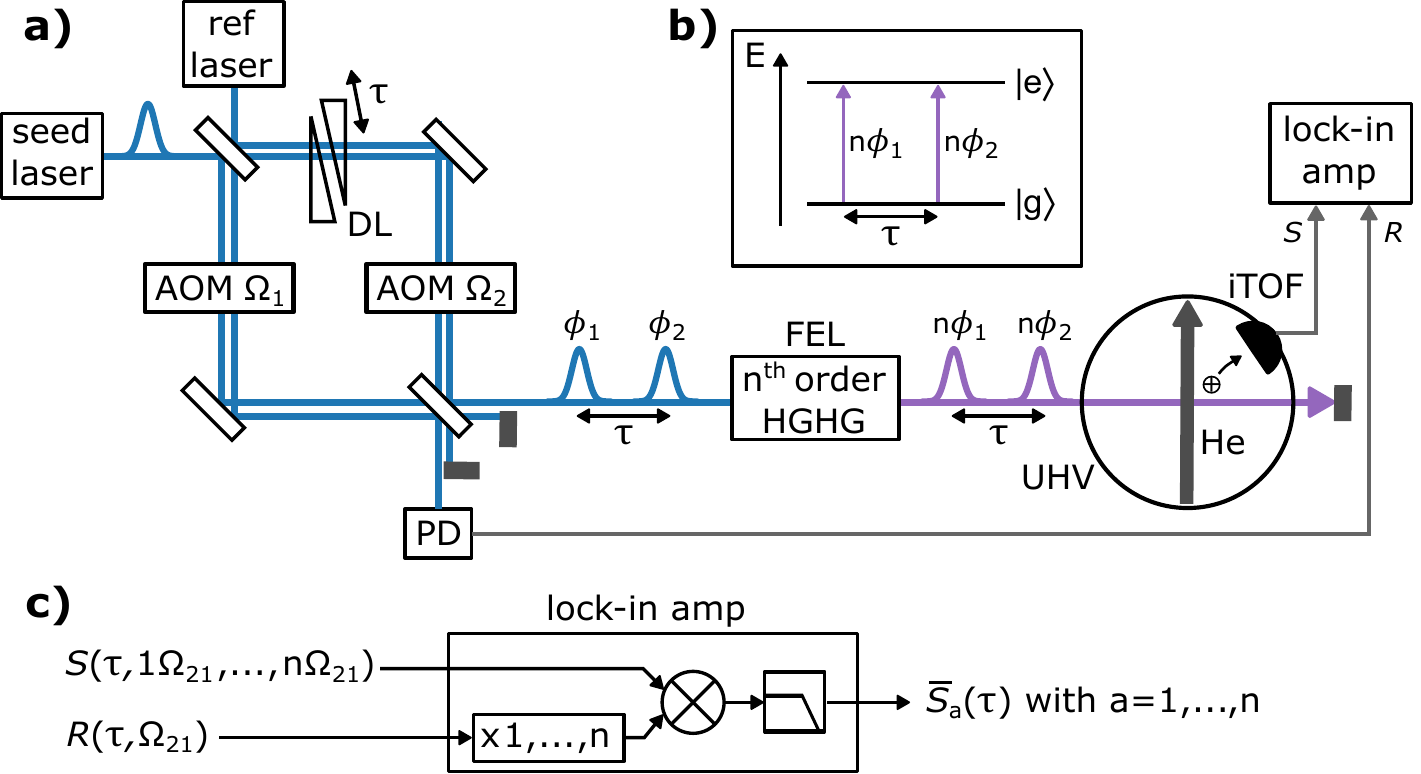}
\caption{Experimental Setup. a) UV seed laser pulses are generated in a monolithic interferometric setup. 
The delay is controlled with a delay line (DL) consisting of two movable glass wedges, the phase $\phi_{i} =\Omega_i t$ is controlled using acousto-optical modulators (AOMs) driven at distinct radio frequencies $\Omega_i$ in phase-locked mode. 
The interferometer is monitored with a continuous wave reference laser, whose interference is recorded with a photodiode (PD) and serves as a reference ($R$) for lock-in detection. 
The pulse pair seeds the high-gain harmonic generation process (HGHG) in the free-electron laser (FEL). 
The resulting XUV pulse pair enters the experimental chamber, which is under ultra-high vacuum (UHV).
Ions created by IR ionization of the excited He (not shown) are detected with an ion time-of-flight spectrometer (iTOF). 
The signal ($S$) from the detector is mass gated, amplified and fed into the lock-in amplifier.
b) Interaction of a two-level system with two phase-modulated FEL pulses.
c) Principle of harmonic lock-in detection. 
Harmonics of the reference signal $R$ are created in the internal electronics of the lock-in amplifier. 
The signal $S$ is demodulated with this reference waveform and contributions at a specific harmonic of the modulation frequency $\Omega_{21} = \Omega_2 -\Omega_1$ are efficiently isolated from the total ion signal.}
\label{fig:setup}
\end{figure}

\section{Experimental method}

\subsection{Phase-modulated wave-packet interferometry}\label{sec:PMWPI}
We use the technique of phase-modulated WPI to track the coherent evolution of XUV electronic wave packets in helium atoms.
Here only a brief description is presented. 
More details about the technique and its extension to the XUV regime can be found in Refs. \citenum{wituschek_tracking_2020, tekavec_wave_2006}, and \citenum{bruder_phase-modulated_2017}.
A phase-locked XUV pulse pair is created by twin-pulse seeding of the HGHG process at FERMI [see Fig.\,\ref{fig:setup}\,a)].  
Each pulse prepares an electronic wave packet: a coherent superposition between the electronic ground state $|g\rangle$ and an excited state $|e\rangle$ [see Fig.\,\ref{fig:setup}\,b)].
The interference of the wave packets is recorded as a function of the inter pulse delay $\tau$ by monitoring the excited state population.
We imprint a low frequency beat $\Omega_{21} = \Omega_{2}-\Omega_{1}$ of a few Hz on the relative phase $\phi_{21} = \Omega_{21} t$ of the seed pulses, using two acousto-optical modulators (AOMs) driven at distinct, phase-locked radio frequencies $\Omega_i.$
Note that here $t$ represents the laboratory time frame.
This leads to a $n\Omega_{21}t$-modulation of the relative phase of the $n^\text{th}$ order FEL harmonic\,\cite{wituschek_tracking_2020} and consequently to a modulation in the WPI signal in the laboratory time frame:
\begin{equation}\label{eq:sig}
S(\tau,t) \propto 1 + \cos(\omega_\text{eg}\tau + n\Omega_{21} t),
\end{equation}
where $\omega_\text{eg} \approx n\omega_0$ denotes the XUV transition frequency from the ground to the excited state, and $\omega_0  \approx\,4.75\,\mathrm{eV}/\hbar$ is the seed laser angular frequency.

At the same time, the seed interferometer is monitored with a single-frequency 266\,nm continuous-wave reference laser, exhibiting the same phase modulation as the seed pulses and tracing all timing and phase jitter inside the interferometer. 
This is used to create an electronic reference signal, whose $m^\text{th}$ harmonic is created in the electronic circuits of the lock-in amplifier
\begin{equation}\label{eq:ref}
R(\tau,t,m) \propto \cos\left[m(\omega_\text{ref}\tau + \Omega_{21} t)\right],
\end{equation}
where $\omega_\text{ref}$ denotes the reference laser angular frequency and $m =1,2,\dots$ is the harmonic demodulation order.
By referencing a lock-in amplifier to the reference signal $R(\tau,t,m=n)$, we can isolate the quantum interference signal and obtain the demodulated WPI signal
\begin{equation}
\bar{S}(\tau) \propto \cos(n(\omega_\text{eg}-\omega_\text{ref})\tau).
\end{equation}
The process of harmonic lock-in detection is schematically depicted in Fig.\,\ref{fig:setup}\,c).
Due to demodulation with the optically generated reference, the majority of the jitter originating from the interferometer cancels.
Furthermore, the fast oscillation frequencies are downshifted to $\overline\omega_\text{eg} = n(\omega_\text{eg}-\omega_\text{ref})$, a technique called rotating-frame detection. 
In this work the oscillation frequencies are reduced by a factor of 50 - 100.
This is particularly beneficial for the XUV spectral regime, as the electronic oscillation frequencies scale directly with the photo-excitation energy. 
At the same time, the impact of timing or phase jitter in the interferometer scales with $\overline\omega_\text{eg}$ instead of $\omega_\text{eg}$, which significantly relaxes demands on interferometric stability in the setup. Finally, the number of data points to sample the oscillation decreases drastically, which is beneficial whenever the total measurement time is critical. 
Throughout this manuscript we will label the different harmonic lock-in demodulation channels with $1\omega_0$, $2\omega_0$,\dots, keeping in mind that $\omega_0$ is the seed laser angular frequency.

\subsection{Experimental setup}
The experimental setup is shown in Fig.\,\ref{fig:setup}\,a).
The seed laser consists of a frequency-tripled, amplified Ti:Sapphire laser system providing the following pulse parameters: central wavelength $\lambda=$ 261\,nm, pulse duration $\Delta t=$ 100\,fs FWHM, repetition rate 50\,Hz.
The UV pulse pair (E\textsubscript{pulse} = 5-10\,\textmu J per pulse) is generated in a highly stable Mach-Zehnder interferometer\,\cite{wituschek_stable_2019}. 
We control the relative delay $\tau$ between the pulses using a pair of movable fused-silica wedges. 
The relative phase $\phi_{21}$ between the pulses is controlled with the AOMs.
A continuous wave reference laser is superimposed with the seed laser in the optical interferometer.
Its interference is recorded using a photo-diode and used as the reference signal for lock-in demodulation.

The pulse pair propagates to the modulator section of the FEL, where it seeds the $n$\textsuperscript{th}-order HGHG process ($n=5,6$ in this work). 
In the regime of temporally well-separated seed pulses ($\tau \geq \text{1.5} \Delta t =$\,150\,fs), we obtain a clean XUV pulse pair at the $n$\textsuperscript{th} harmonic with very precise delay $\tau$ and relative phase $n\phi_{21}$\,\cite{wituschek_tracking_2020}.
In this case, an upper delay limit $\tau\leq 1.3$\,ps is given by the finite length of the electron bunch used for HGHG.
When the seed pulses overlap temporally ($|\tau| <150$\,fs), the situation is different. 
Here, the nonlinear HGHG process introduces a mixing between contributions of either of the two seed pulse electric fields. 
This effect is discussed in detail in sections\,\ref{sec:HHAC} and \ref{sec:WPI}.
The FEL was operated at 50\,Hz, in a regime of weak electron beam-modulation, where the XUV harmonic pulse-duration is expected to be proportional to $\Delta t/\sqrt{n}$\,\cite{finetti_pulse_2017}.

Experiments were carried out at the Low-Density Matter (LDM) endstation at FERMI\,\cite{lyamayev_modular_2013}.
A helium beam was generated by supersonic expansion through a pulsed nozzle, yielding a density of $\approx$\,10\textsuperscript{13}\,cm\textsuperscript{-3} in the interaction region of the ion time-of-flight mass spectrometer.
The focus diameter of the FEL in the interaction volume was 70\,\textmu m FWHM.
A synchronized near-IR fs laser, delayed by $\approx 5$\,ps was used to ionize the excited helium in the WPI experiments while we ionized helium directly in a one-photon process for the autocorrelation (AC) measurements using a FEL energy above the He ionization potential. 
The time-of-flight gated ion yields were amplified and fed into the lock-in amplifier for processing.

\section{High order interferometric autocorrelation}\label{sec:HHAC}
Within the temporal overlap region of the seed pulses ($\vert\tau\vert\leq 1.5\Delta t \approx$  150\,fs) interference occurs, leading to a modulation of the seed pulse intensity from zero to four times the intensity of a single seed pulse.
The nonlinear response of the HGHG process to this intensity modulation leads to a considerable modification in the spectrotemporal properties of the resulting FEL pulses.
To understand this behaviour we measured linear, fringe-resolved FEL ACs at a specific harmonic (here the sixth harmonic).
Assuming that the HGHG process acts as an ideal sixth order nonlinear process, the linear FEL AC corresponds to the sixth order AC of the seed laser pulses, which is straight-forward to calculate.
Comparison of the measured ACs to calculations provides valuable insights into the nature of the HGHG process itself, when seeded with temporally overlapping pulses.
A similar study using second-harmonic generation in nonlinear cystals has been previously reported in Ref.\,\cite{bruder_phase-modulated_2017}.

This section is structured as follows: in Sec.\,\ref{sec:no_saturation_AC} we calculate the ideal sixth order interferometric AC of the seed pulses and show how its individual harmonic components can be isolated using the phase-modulation technique. In Sec.\,\ref{sec:results_AC} we present the experimental results of the AC measurement and show that they differ significantly from the ideal sixth order AC.  Finally, in Sec.\,\ref{sec:saturation_AC} we introduce a more realistic model for the response of the HGHG process to the seed pulse intensity-modulation and compare this model to our experimental data.
For the remainder of this manuscript we will refer to the ideal AC model from Sec.\,\ref{sec:no_saturation_AC} as the `unsaturated' case, while we refer to the model from Sec.\,\ref{sec:saturation_AC} as the `saturated' case.

\subsection{Ideal high-order interferometric autocorrelation}\label{sec:no_saturation_AC}
\begin{figure}
\centering\includegraphics[width=0.65\linewidth]{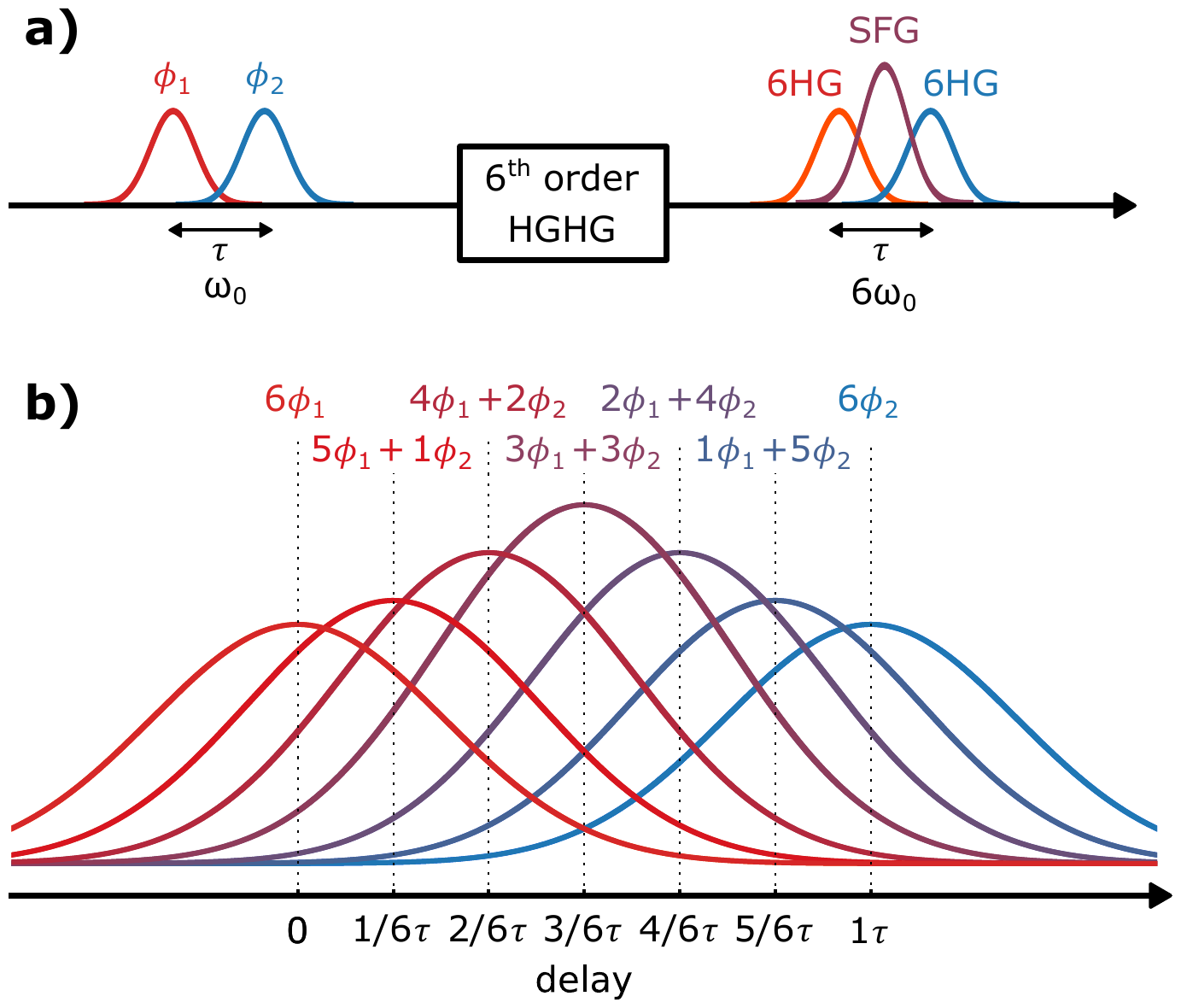}
\caption{Harmonic generation with temporally overlapping seed-pulse pairs.
a) During the sixth-order HGHG process the XUV radiation at frequency $6\omega_0$ can be created by sum-frequency generation (SFG) of different fractions of either of the seed pulses, or by direct sixth-harmonic generation (6HG) of the individual pulses.
b) Close-up of the components of the resulting XUV field. 
The individual 6HG and SFG contributions to the total XUV field are modulated with different modulation frequencies and separated in time by increments of  $\tau /6$ (see also Eq.\,\ref{eq:AC3}). 
Note that the seed electric field is oscillating at the seed-laser frequency $\omega_0$ and all components of the XUV field are oscillating at the frequency $6\omega_0.$ 
We use the red and blue colors to distinguish the different modulation frequencies.
 }
\label{fig:ac6hsfg}
\end{figure}

We define the electric field of the seed pulse:
\begin{equation}
E_i(t)  =  A(t,\Delta t) \cos \left[ \left( \omega_0 + \Omega_i \right) t\right] \label{eq:AC0}
\end{equation}
where $i=$~1,2 denotes the branch of the interferometer, $\omega_0 =$ 4.75\,eV$/\hbar$ is the seed laser angular frequency and $A(t,\Delta t)$ is a Gaussian amplitude function with a FWHM of $\Delta t.$  
The sixth-order interferometric AC is then given by:

\begin{equation}
AC(\tau)= \int \left\vert\left(E_1(t)+E_2(t-\tau)\right)^6\right\vert^2 \mathrm{d} t.\label{eq:AC1}
\end{equation}
The integrand in Eq.\,\ref{eq:AC1} can be expanded after writing

\begin{equation}
\left(E_1(t)+E_2(t-\tau)\right)^6 = \sum_{k=0}^6 \binom{6}{k} E_1(t)^{6-k} E_2(t-\tau)^k.\label{eq:AC2}
\end{equation}
With Eq.\,\ref{eq:AC0} we can describe each summand of Eq.\,\ref{eq:AC2} and obtain:
\begin{align}\label{eq:AC3}
E_1(t)^{6-k} E_2(t-\tau)^k =&~A\left(t-k\tau/6,\Delta t/\sqrt{6}\right)\,
A\left(\sqrt{k(6-k)}\tau,\sqrt{6}\Delta t\right) \\
& \times \cos\left[ 6\omega_0 \left(t-k\tau/6\right) + \left((6-k)\Omega_1 + k\Omega_2\right)t\right] \nonumber
\end{align}

The cases where $k=$\,0 or $k=$\,6 correspond to direct sixth-harmonic generation (6HG) of the individual pulses in the pulse pair. 
For all other values of $k$ we obtain sum-frequency generation (SFG), meaning that the resulting XUV electric field consists of a mixture of components from both pulses [see Fig.\,\ref{fig:ac6hsfg}\,a)].
The first term in Eq. \ref{eq:AC3} shows that the SFG pulse is shifted in time by increments of $k\tau/6.$ 
The second term describes the decreasing amplitude of the SFG pulses with increasing delay, indicating that they only occur when the seed pulses overlap temporally.
The last term shows that each component has a distinct phase signature $((6-k)\Omega_1 + k\Omega_2)t.$
This situation is shown in Fig.\,\ref{fig:ac6hsfg}\,b).

Pairwise interference of the summands in Eq.\,\ref{eq:AC2} gives rise to the individual contributions to the total AC signal $AC(\tau)$, ranging from $1\omega_0$ to $6\omega_0$ interference components. 
As an example, we calculate one contribution explicitly: the interference between $E_1(t)^6$ and $E_1(t)^4 E_2(t-\tau)^2$, neglecting envelope terms in Eq.\,\ref{eq:AC3}:
\begin{equation}\label{eq:AC2H}
AC_{2\omega_0}(t,\tau) \propto
\left\vert E_1(t)^6 + E_1(t)^4 E_2(t-\tau)^2 \right\vert^2 \propto 
1 +  \cos\left[ 2\omega_0 \tau - 2\Omega_{21} t \right]
\end{equation}
This $2\omega_{0}$-component is modulated with a frequency of $2\Omega_{21}$ and oscillates at a frequency of $2\omega_0$ with respect to $\tau.$ 
All other components can be derived in analogy and their modulation frequencies are summarized in Tab.\,1.

Due to its distinct modulation frequency we can isolate the $AC_{2\omega_0}$-component in Eq.\,\ref{eq:AC2H} from the total signal by second-harmonic lock-in detection with the reference $R(\tau,t,m=2).$ This yields:
\begin{equation}\label{eq:AC2H_2}
\overline{AC}_{2\omega_0}(\tau) \propto \cos\left[ 2(\omega_0-\omega_\text{ref})\tau \right]
\end{equation}
Note, that the fast quantum interferences at a frequency of $2\omega_0$ are downshifted by $2\omega_\text{ref}$ due to the rotating-frame detection.

The calculated the AC traces at each harmonic $m\omega_0$ ($m=1,\dots,6$) of the fundamental frequency are shown in Fig.\,\ref{fig:ac6h}\,a).
The individual $m\omega_0$ components comprise of a Gaussian envelope and are oscillating at a frequency of $m(\omega_0-\omega_\text{ref})$.

\begin{table}
\centering
  \begin{tabular}{ c | c || c | c}
	AC component & Mod. freq. & AC component & Mod. freq. \\ 
    \hline \hline
    $|E_i^6 + E_i^5 E_j|^2$ & $1\Omega_{21}$ & $|E_i^5 E_j + E_i^4 E_j^2|^2$ & $1\Omega_{21}$ \\ 
    $|E_i^6 + E_i^4 E_j^2|^2$ & $2\Omega_{21}$ & $|E_i^5 E_j + E_i^3 E_j^3|^2$ & $2\Omega_{21}$ \\     
    $|E_i^6 + E_i^3 E_j^3|^2$ & $3\Omega_{21}$ & $|E_i^5 E_j + E_i^2 E_j^4|^2$ & $3\Omega_{21}$ \\    
    $|E_i^6 + E_i^2 E_j^4|^2$ & $4\Omega_{21}$ & $|E_i^5 E_j + E_i E_j^5|^2$ & $4\Omega_{21}$ \\     
    $|E_i^6 + E_i^1 E_j^5|^2$ & $5\Omega_{21}$ & $|E_i^4 E_j^2 + E_i^3 E_j^3|^2$ & $1\Omega_{21}$ \\     
    $|E_i^6 + E_j^6|^2$ & $6\Omega_{21}$ & $|E_i^4 E_j^2 + E_i^2 E_j^4|^2$ & $2\Omega_{21}$ \\

  \end{tabular}
  \label{tab:components}
  \caption{Summary of all components contributing to the AC signal and their respective modulation frequency. It is $i,j =1,2$ with $i\neq j.$ For simplicity we omit all coefficients arising from the exponentiation in Eqs.\,\ref{eq:AC1}\,\&\,\ref{eq:AC2}.}
\end{table}

\subsection{Experimental results}\label{sec:results_AC}
In the experiment, the linear interferometric AC of the sixth harmonic FEL pulses was recorded.
To this end, He atoms were ionized ($\text{IP} = 24.6$\,eV) with the sixth FEL harmonic (28.5\,eV) in a one photon process.
The FEL intensity was reduced using metal filters to avoid saturation in the ionization yields and in the detection electronics.
The AC was recorded for delays ranging from -200 to 200\,fs.
In analogy to second-harmonic interferometric ACs where the DC, $1\omega_0$ and $2\omega_0$ contributions can be isolated by Fourier-filtering\,\cite{wollenhaupt_femtosecond_2007}, the $1\omega_0$-$6\omega_0$-contributions were isolated by harmonic lock-in detection.
The demodulated signal for the individual harmonic contributions is shown in Fig.\,\ref{fig:ac6h}\,c).
The disagreement between the experimental data and the calculations of the ideal AC in the previous section is clearly observable.
In order to understand the experimental results we performed numerical simulations based on high-order interferometric ACs including a simplified model of the XUV generation in the HGHG process, which is described in the following section.

\subsection{FEL amplification and saturation in HGHG}\label{sec:saturation_AC}
In this section we analyze the modifications occurring to the field envelope during the harmonic conversion and the following FEL amplification processes, including the correction due to the onset of saturation.
We exclude from our analysis short pulse effects and deep saturation effects, which would modify the dispersion during the interaction process and could not be modeled in terms of field amplitudes\,\cite{yang_postsaturation_2020}.
\begin{figure}
\centering\includegraphics[width=1.0\linewidth]{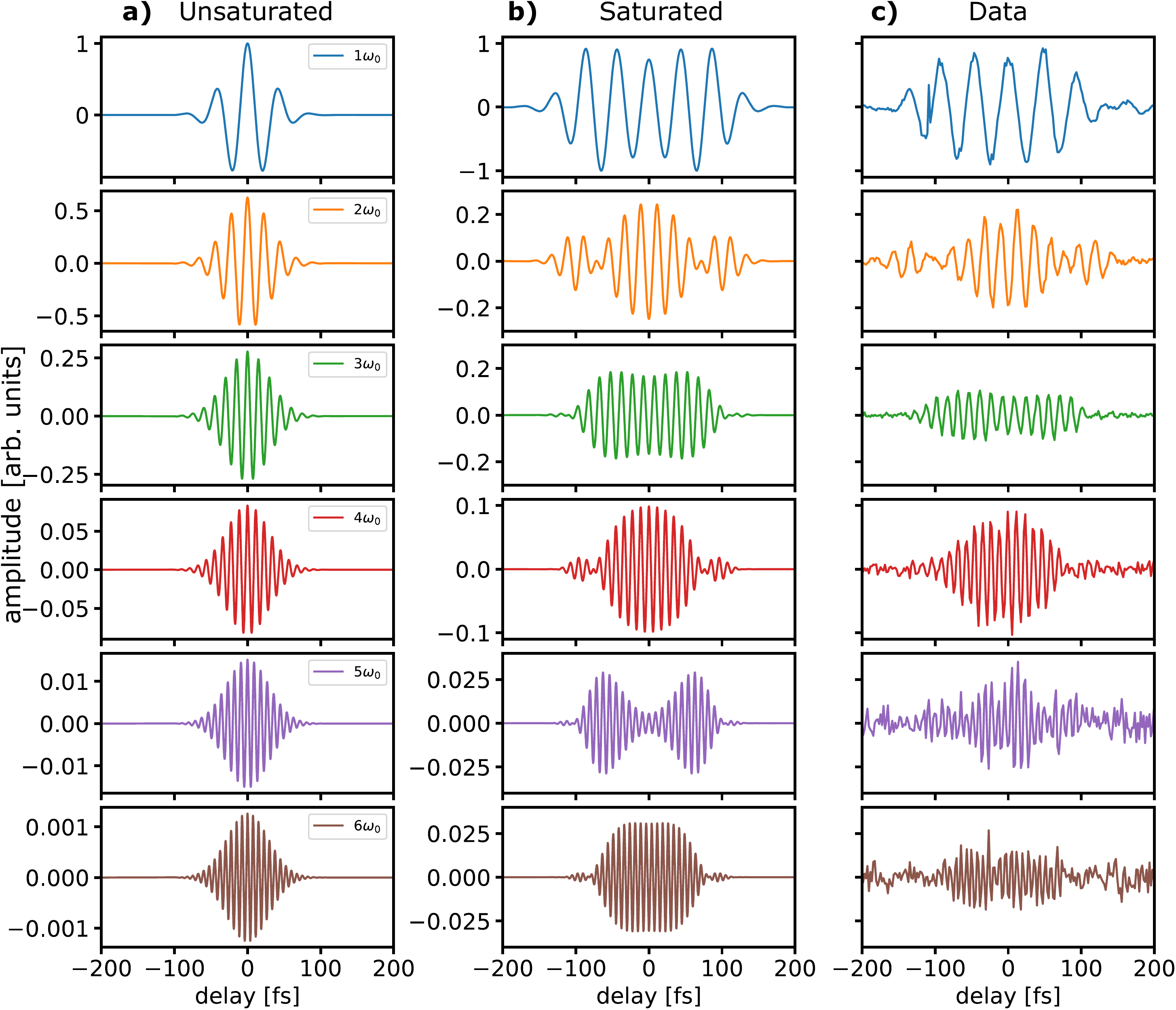}
\caption{Sixth-order interferometric AC. 
Calculation of the $1\omega_0$-$6\omega_0$ components of the 6th-order fringe-resolved AC for a) the unsaturated case of Eq.\,\ref{eq:AC1} and b) for the case with the saturation (Eq.\,\ref{eq:ACsat}).
The experimental data as obtained from harmonic lock-in detection is plotted in c).
Note the different range on the y-axes, b) and c) share the same y-axis. 
Due to rotating-frame detection the interference fringes appear at much smaller oscillation frequency ($\approx 8$\,fs instead of 145\,as, for the $6\omega_0$-component). }
\label{fig:ac6h}
\end{figure}

In HGHG, the coherent bunching produced through the interaction of the electron beam with the external seed laser\,\cite{yu_generation_1991} is further amplified by the FEL process occurring in the long radiator.
The FEL output pulse inherits the spectrotemporal properties of the seed
pulse\,\cite{gauthier_spectrotemporal_2015}. 
These properties are preserved when the FEL is operated at low seed intensity and far from saturation. 
In case of an excess of seed laser field or of a strong amplification in the radiator, the FEL amplification process distorts the initial amplitude of the seed pulse. 
At the modulation stage, an excess of seed laser electric field leads to overbunching of the electrons, folding the longitudinal phase space and inducing modifications in the spectrotemporal pattern of the resulting FEL pulse\,\cite{gauthier_spectrotemporal_2015, de_ninno_chirped_2013}.
Furthermore, a large seed induces a strong energy modulation and bunching, and anticipates the saturation process in the amplifier.
In the temporal overlap region, the seed-laser electric field amplitude is modulated from zero to twice its initial value, assuming an ideal interference contrast.
Therefore the corresponding seed power in the overlap region can be up to four times larger than the peak power of the single pulse. This has to be carefully taken into account both at the bunching stage and at the amplification stage, to predict the output FEL spectrotemporal
properties. 

The number of coherent electrons produced as a result of the interaction with the seed in the modulator, after the following dispersive section is determined by the bunching factor
\begin{equation}
b(t)=e^{-\frac{1}{2}({n}{B})^{2}}\times J_{n}({n}{A(t)}{B})\label{eq:hghgBUNCH}
\end{equation}
where $J_{n}$ is the $n^\text{th}$ order Bessel function and $n$ is the FEL harmonic. $B=k R_{56} \sigma_E$ is proportional to the seed wavevector $k$ and the strength of the dispersion $R_{56}$, which is used to convert the energy modulation of the electron beam into a density modulation. $\sigma_E$ is the relative energy spread of the electron beam and $A(t) =\Delta_E(t)/\sigma_E$, where $\Delta_E(t)$ is the energy modulation produced by the seed laser, and is proportional to the seed laser field $E_\text{tot}(t).$
In the case of double-pulse seeding $E_\text{tot}$ is the superposition of the electric field of the two seeds
\begin{equation}
E_\text{tot}(t,\tau)=E_{1}(t)+E_{2}(t-\tau), \label{eq:Etot}
\end{equation}
that also depends on the delay $\tau$ between the two seed laser pulses.

The FEL electric field amplitude emitted in the first part of the radiator is directly proportional to the bunching $b(t)$ 
\begin{equation}
E_\text{FEL}(t)\propto b(t).\label{eq:bunch}
\end{equation}
Note that if the FEL is operated at low dispersion $B$ or at low seed-intensity, the argument of the Bessel function is small and we may expand it in a Taylor series, obtaining at the lowest order 
\begin{equation}
E_\text{FEL}(t,\tau)\propto E_\text{tot}(t,\tau)^{n}.\label{eq:expo}
\end{equation}

\begin{figure}
\centering\includegraphics[width=0.7\linewidth]{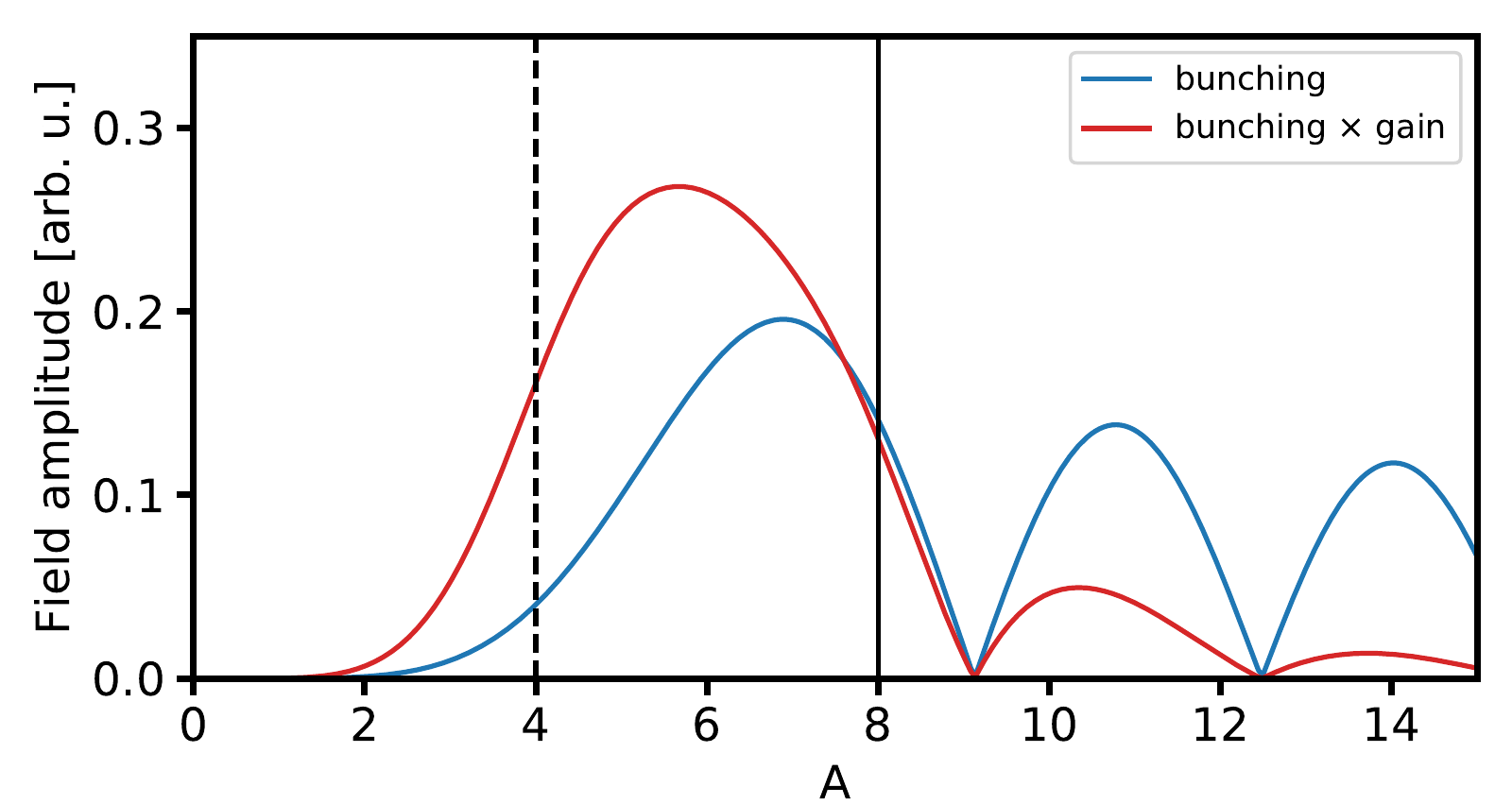}
\caption{ a) Functions used to model the FEL amplitude. HGHG bunching (Eq.\ref{eq:hghgBUNCH}, blue)
and bunching plus gain and saturation (Eq.\ref{eq:Bunchgain}, 
red). The harmonic order is $n=6$ in this graph.
The vertical black lines indicate the maximum values of $A$ used in the simulation shown in Fig.\,\ref{fig:ac6h}\,b). Here the dashed and solid lines indicate the maximum value for $A$ for a single seed pulse and for the two seed pulses which interfere constructively at $\tau = 0$, respectively.
}
\label{fig:bessel_sigmoid} 
\end{figure}

If instead the field and/or the dispersion are large enough to maximize the bunching factor, the Bessel function induces a non-linear transformation of the seed amplitude as given by Eq.\,\ref{eq:hghgBUNCH}.
In a realistic scenario, also the modifications induced by the FEL amplification process in the radiator need to be considered\,\cite{finetti_pulse_2017, giannessi_seeding_2020}.
The FEL power grows in the amplifier as
\begin{equation}
P_\text{FEL}(t,\tau,z)\propto b(t,\tau)^{2}e^{z/L_{g}}
\end{equation}
where $z$ is the propagation direction along the radiator and $L_{g}$ is the power folding gain length\,\cite{bonifacio_collective_1984}.
The dependence of the gain length on realistic electron beam and undulator parameters was investigated by Xie\,\cite{xie_design_1995}.
An estimate based on the Xie model in the experimental conditions of FERMI, assuming a natural electron-beam energy spread of 140\,keV, provides $L_{g}\approx 2\,\mathrm{m}.$
The gain length has a strong dependence on the electron beam energy spread: a larger energy spread corresponds to a longer gain length.
The time-dependent energy modulation caused by the seed, represents an additional energy-spread contribution, which adds in quadrature to the natural electron-beam energy spread $\sigma_{E}.$
The bunching process itself induces therefore a time dependent growth of the gain length $L_{g}\rightarrow L_{g}\chi(t).$
An expression for this dependence was deduced from the Xie scaling relations, which in the specific FERMI operating conditions is
\begin{equation}
\chi\left(L_{g},\sigma_{E},A(t)\right)=\frac{1+3\left(4\pi L_{g}\sigma_{E}/\lambda_{u}\right)^{2}\left(1+A(t)^{2}+0.01 A(t)^{4}\right)}
{1+3\left(4\pi L_{g}\sigma_{E}/\lambda_{u}\right)^{2}},
\end{equation}
where $\lambda_u$ is the undulator magnet period.
This correction affects both the gain length and the final saturation power, which is reduced proportionally to $\chi^{-2}.$ 
The FEL field at the end of the radiator of length $L_{u}$ is obtained by approximating the FEL power saturation with a logistic function\,\cite{dattoli_semi-analytical_2002}.
We therefore obtain
\begin{equation}
E_\text{FEL}(t,\tau) \propto \frac{1}{\chi(t,\tau)}\left[\frac{\frac{1}{2}b^2(t,\tau)e^{\frac{L_u}{L_{g}\chi(t,\tau)}}}{1+\frac{1}{2}b^2(t,\tau)e^{\frac{L_u}{L_{g}\chi(t,\tau)}}}\right]^{1/2}.\label{eq:Bunchgain}
\end{equation}
In Fig.\,\ref{fig:bessel_sigmoid} we show a comparison between different approximations for the FEL electric field amplitudes for the case of a single seed pulse and $n=6.$

The following equation is used for numerical calculation of the AC signals for the saturated case:
\begin{eqnarray}
AC_\text{sat}(\tau) & = & \int \vert E_\text{FEL}(t,\tau) \vert^2 \mathrm{d} t. \label{eq:ACsat}
\end{eqnarray}

The results of the calculations for the saturated case are shown in Fig.\,\ref{fig:ac6h}\,b).
The seed-laser electric-field amplitude was used as a free parameter in the calculations. 
In the shown calculations the amplitude of a single seed pulse was such that $\max_{t} \vert A(t)\vert\approx 4$ (cf. dashed black line in Fig.\,\ref{fig:bessel_sigmoid}).
One can clearly see a broadening of the AC trace and drastic changes in amplitude and phase compared to the non-saturated calculations in Fig.\,\ref{fig:ac6h}\,a).
Both effects are a direct consequence of saturation in the harmonic generation.
The calculations for the saturated case agree well with the data. 
The correct amplitudes are reproduced for all orders $1\omega_0$-$6\omega_0.$ 
Also the phase of the fringes is reproduced well for $1\omega_0$-$4\omega_0.$ 
However, for $5\omega_0,6\omega_0$ the noisy data prevents a quantitative comparison with the calculations.
Note, that calculations which take only into account the bunching factor (cf. Eq.\,\ref{eq:bunch}), and disregard the influence of gain saturation have a poor agreement with the data (not shown).
The slight asymmetry in the data is attributed to inhomogeneities present in the electron bunch used for the HGHG\,\cite{mahieu_two-colour_2013}.
Here the pulse which is scanned along the electron-bunch exhibits slightly different conditions at each position, leading to delay dependent spectrotemporal properties.

Note that no calibration of the phase-transfer function of the electronics (detectors, amplifiers, etc.) in the electronic signal pathways for $S$ and $R$ was performed.
Frequency-dependent phase-transfer functions can lead to an individual offset of the absolute phase of the harmonic AC contributions. 
For better comparison of data and calculation we set the fringe phase $\phi_\text{data}(\tau =0) \equiv \phi_\text{calc}(\tau=0)$ for each harmonic in Fig.\,\ref{fig:ac6h}.

\section{Wave-packet interferometry in helium}\label{sec:WPI}
\begin{figure}
\centering\includegraphics[width=0.8\linewidth]{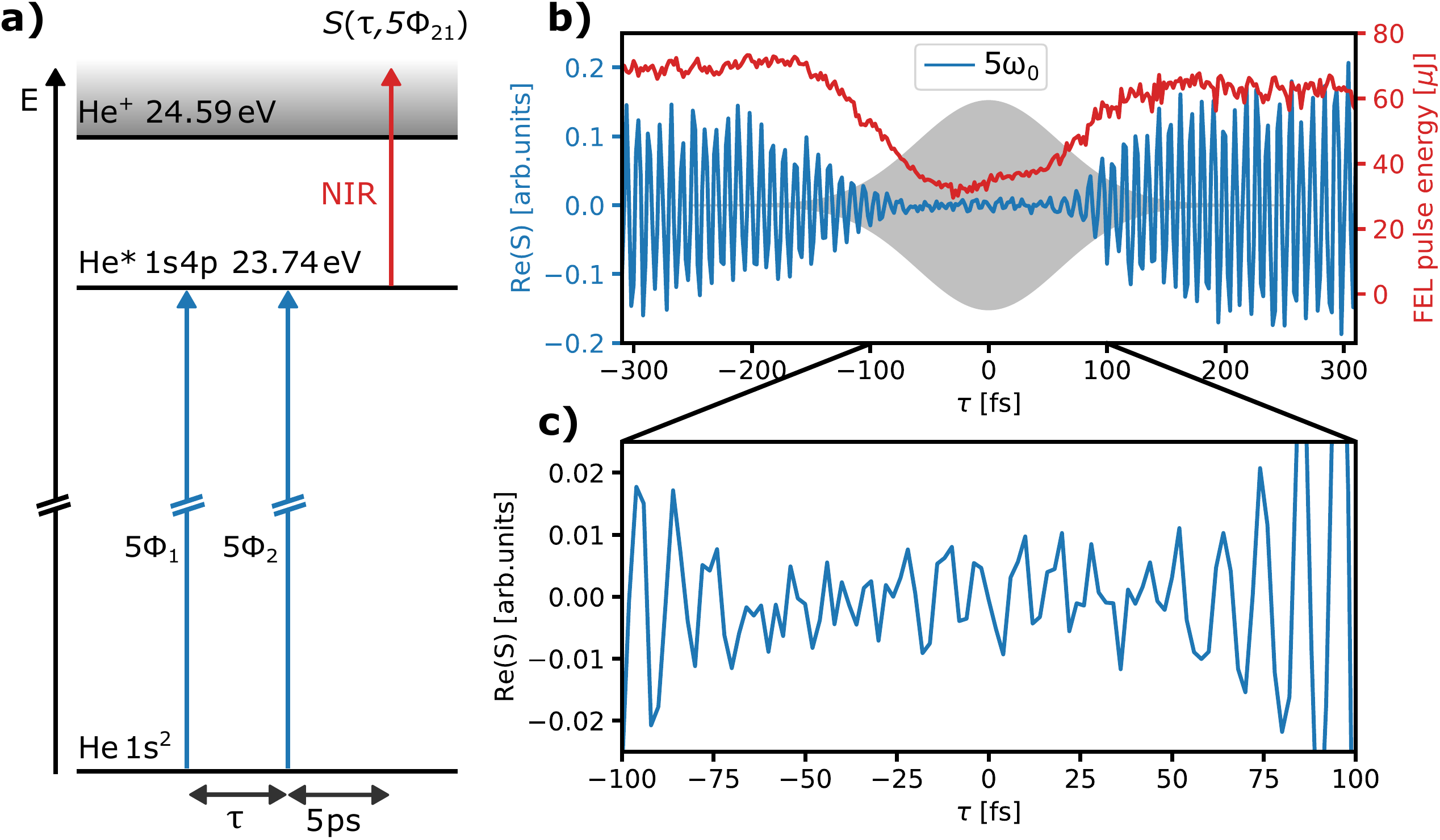}
\caption{Wave-packet interferometry in helium. a) Level scheme with all relevant transitions and laser-pulse interactions. b) Time-domain interferogram showing the decay of the WPI signal in the temporal overlap region (blue). Note that the fast 174\,as-period quantum interferences are downshifted to $\approx 9$\,fs by rotating-frame detection. The grey shaded area depicts the calculated linear AC of the seed pulses. FEL pulse energy averaged over 640 FEL shots (red). The impact of saturation in the overlap region is clearly visible. c) Zoom into the overlap region.}
\label{fig:helium}
\end{figure}

In the previous section, we studied the impact of saturation on off-resonant one-photon ionization of helium by performing fringe-resolved AC measurements. 
We now focus on the resonant excitation of helium and study the impact of saturation on the WPI measurements.
Here, helium served as a two-level system and we probed the evolution of electronic wave packets between the ground state $\vert g\rangle = \vert 1s^2\rangle$ and the excited state $\vert e\rangle = \vert 1s4p\rangle$ [Fig.\,\ref{fig:helium}\,a)].
Note that this coherence spans over 23.74\,eV, corresponding to an oscillation period of only $T = h/E = 174$\,as. 
Due to the rotating-frame detection this fast oscillation is downshifted to $\overline{T} \approx 9$\,fs period. 
We tuned the FEL to the $n=5$ harmonic, to be resonant to the transition. 
At the same time, no other resonances overlapped with the spectrum of the FEL.
The pulse-energy was reduced to 30\,nJ per pulse (using metal filters) to avoid saturation of the transition.
We used a synchronized near infrared (NIR) laser delayed by $\approx$\,5\,ps to ionize selectively the He atoms in the excited state, hence the measured photoproduct yields reflect the excited state population.

We scanned the delay from -300\,fs to 300\,fs in 2\,fs increments, averaging over 640 FEL shots for each delay step. 
The time-domain interferogram obtained by fifth-harmonic lock-in detection is shown in Fig.\,\ref{fig:helium}\,b).
Far away from the overlap region, the wave-packet signal has constant amplitude (cf.~\cite{wituschek_tracking_2020}). 
This is consistent with the nanosecond lifetime of the $4p$-state in isolated helium atoms.
As the delay is reduced to zero, the amplitude of the oscillation decreases and only very weak oscillations remain below $\vert\tau\vert\leq 60$\,fs.
At the same time, the FEL pulse energy, averaged over multiple $\Omega_{21}$-modulation periods, drops significantly due to saturation [Fig.\,\ref{fig:helium}\,b)].
Even in the pulse-overlap region, we can still observe oscillatory WP signals, which is surprising, considering the reduced signal amplitude by a factor of $\approx 10$ [Fig.\,\ref{fig:helium}\,b)].
Note that the pulse energy and the central wavelength of the pulse which is scanned along the electron bunch are changing slightly as a function of $\tau$, due to the non-uniform energy and density profile of the electron bunch\,\cite{gauthier_spectrotemporal_2015}.
The resulting variation of the overlap with the atomic resonance may explain the higher signal amplitude in Fig.\,\ref{fig:helium}\,b) at positive delays, despite the slightly lower pulse energy in this region.

\begin{figure}
\centering\includegraphics[width=1.0\linewidth]{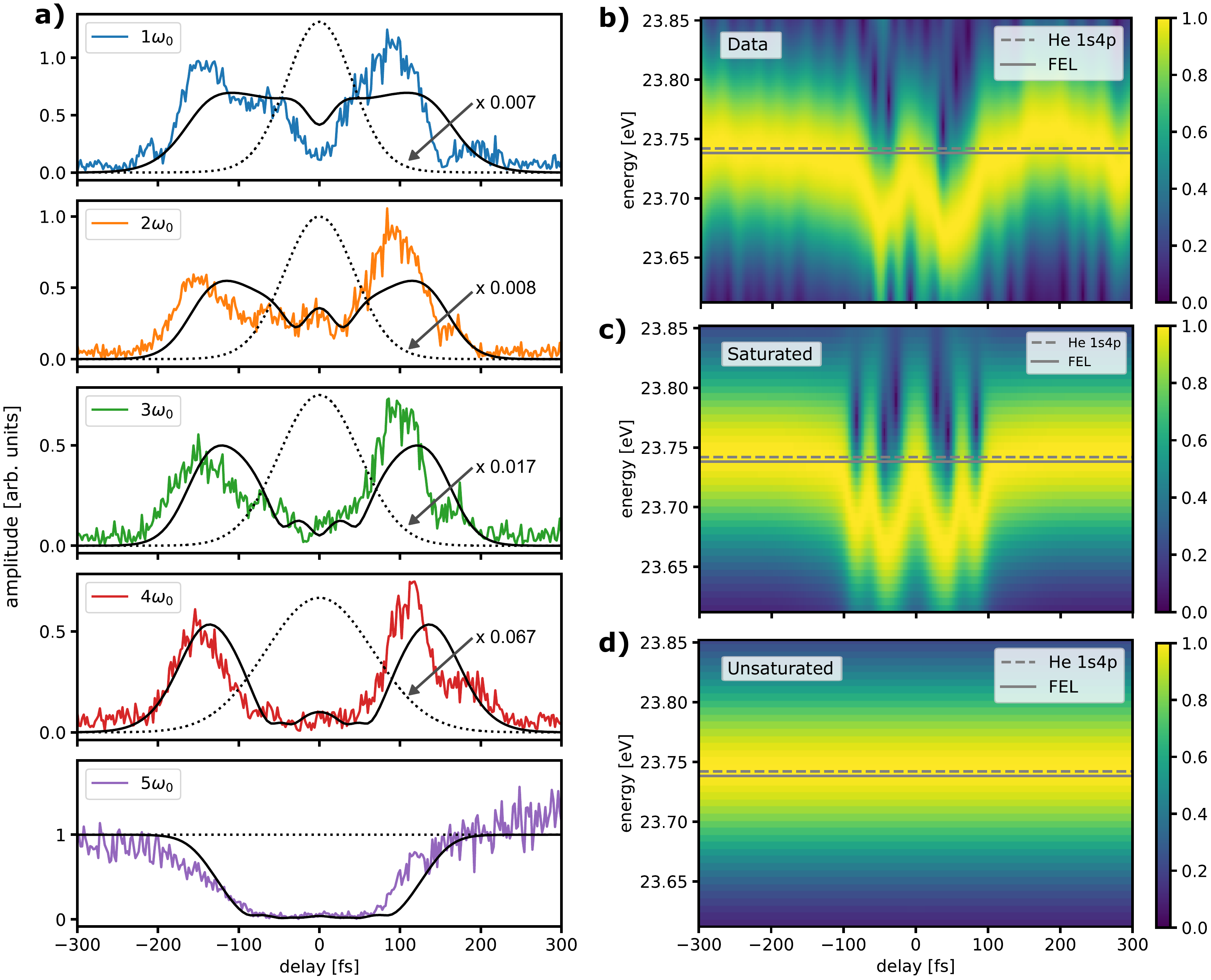}
\caption{a) WPI signal amplitudes obtained for the $1\omega_0$-$5\omega_0$-components (in color). The black curves show the results of our calculations with saturated FEL pulses (solid black lines) and with unsaturated pulses (dotted black lines).
Spectrograms of the $5\omega_0$-components:  b) for the $5\omega_0$-data, c) for the calculated $5\omega_0$-signals with saturation, d) and for the calculated $5\omega_0$-signals without saturation. 
The grey solid and dashed lines indicate the FEL photon energy and the He resonance, respectively. 
Note that for each delay-step of the spectrogram the resulting spectrum was normalized.}
\label{fig:wpi_sim}
\end{figure}

The WPI signal is generated by the interaction of two laser pulses with the helium atom\,\cite{tekavec_wave_2006}. 
In the overlap region the atom can interact with either 5HG or SFG pulses.
In a similar way as in the AC measurement we analyze the various signal contributions using harmonic lock-in demodulation.
The obtained signals are shown in Fig.\,\ref{fig:wpi_sim}\,a). 
Note that in contrast to Fig.\,\ref{fig:helium}\,b) we show only the amplitude of the signal as obtained from the lock-in amplifier.
In the overlap region, one would expect the $5\omega_0$-signal to be a superposition of a high-order (5\textsuperscript{th} order) AC and the regular resonant WPI signal in helium [see also Fig.\,\ref{fig:helium}\,b)]. 
However, this would not explain the transient drop in the $5\omega_0$-signal.
Furthermore, the 1-$4\omega_0$-signals [Fig.\,\ref{fig:wpi_sim}\,a)] do not match with simulated AC signals either.

To understand the signal behavior we calculated the response of the helium two-level system subject to the excitation with unsaturated and saturated FEL double-pulses.
The calculations are based on an exact numerical solution of the Schr\"odinger-equation by a fourth-order Runge-Kutta integrator\,\cite{palacino-gonzalez_theoretical_2017,palacino-gonzalez_theoretical_2017-1}. 
From the calculations the excited state population after interaction with the FEL pulses is obtained.
By imparting a phase-cycling scheme and subsequent Fourier filtering the individual pathways are isolated.

The results of the calculations using saturated FEL pulses are shown as black solid curves in Fig.\,\ref{fig:wpi_sim}\,a).
Alongside, the amplitude of the WPI signal as obtained from the different harmonic demodulation channel is shown. 
This amplitude can be understood as the amplitude of the oscillations seen in Figs.\,\ref{fig:ac6h} and \ref{fig:helium}.
The calculations are normalized to the amplitude of the $5\omega_0$ data in the region $\vert\tau\vert >$\,200\,fs.
The calculations reproduce the data reasonably well, especially for the $5\omega_0$-component, which is the most interesting signal as it contains information on the XUV wave-packet evolution also outside the overlap region.
For the $1\omega_0 - 4\omega_0$-components the major features of the data are reproduced with correct amplitude and timing: two peaks at $\pm$\,150\,fs and a transient decrease for smaller delay values. 
The residual mismatch between data and calculation can be attributed to the simplified FEL harmonic generation model and the fact that chirp, the finite interference contrast of the seed pulses, and inhomogenities of the electron bunch are not included in the calculations.
As a comparison calculations are shown, which involve unsaturated FEL pulses for the interaction with the helium [black dotted curves in Fig.\,\ref{fig:wpi_sim}\,a)].
Note that in this case the amplitude of the $5\omega_0$-component maintains constant amplitude in the overlap region, possibly allowing for WPI studies also in this regime.

In Fig.\,\ref{fig:wpi_sim}\,b) we show the results of a spectrogram analysis sliding a 25\,fs FWHM Gaussian window over the $5\omega_0$ demodulated signal [see Fig.\,\ref{fig:helium}\,b)] in 2\,fs steps. 
In the overlap region the frequency of the interferogram experiences modulations and is generally shifted to the lower energy side of the He resonance.
The frequency shift is much larger than the distance between the resonance and the FEL carrier frequency.
The same spectrogram obtained from our calculations also shows a redshift of the spectral line of the resonance in the overlap region whenever saturation is present [Fig.\,\ref{fig:wpi_sim}\,c)].
This can be understood by taking into account the phase jump of the exciting FEL light which is a consequence of the saturation. For example see the phase jump of the $5\omega_0$-component at $\vert\tau\vert \approx 90$\,fs in Fig.\,\ref{fig:ac6h}\,b). 
This phase jump in the time domain leads to line splitting in the frequency domain.
Our calculations reproduce this effect and show that the spectral amplitude of the split peak is strongly enhanced in the direction of the detuning from the resonance, hence resulting in a shift towards lower energies.

In addition, the same spectrogram analysis was performed using unsaturated FEL pulses [Fig.\,\ref{fig:wpi_sim}\,d)].
Here the WPI frequency is centered at the He resonance frequency throughout the overlap region and no line splitting in the frequency-domain is observed.
This and the constant amplitude of the $5\omega_0$-signal implies, that when operating the FEL below saturation one may be able to extract WPI signals also during temporal overlap of the seed pulses.

%

\section{Discussion and conclusion}
We studied the suitability of the HGHG process for XUV time-domain spectroscopy in the regime where the seed pulses overlap temporally.
To this end, high-order nonlinear interferometric AC and WPI measurements were performed and compared to numerical calculations.
With the help of the AC measurements it was found that the influence of saturation in the HGHG process modifies the obtained interferograms.
However, the good agreement with the calculations indicates that these effects are predictable and that the phase modulation imparted in the seed pulses coherently transfers onto the XUV pulses. 
This shows that the coherence of the HGHG process is conserved even during temporal overlap.
By performing WPI measurements on a well understood two-level model system the region where seed pulse interference affects the time-domain interferograms was confined to $\vert\tau\vert \leq 1.5 \Delta t =: \tau_\text{crit}$, where $\Delta t$ is the FWHM of the seed pulses. 

The calculations indicate that $\tau_\text{crit}$ is not a constant, but depends on the amount of saturation inherent in the harmonic-generation process and reduces to zero as soon as the saturation vanishes. 
Consequently, unperturbed time-domain interferograms can be obtained by reducing the amount of saturation in the HGHG process.
This would allow for unambiguous interpretation of data and therefore may facilitate the extension of XUV time-domain spectroscopy techniques to the region where the seed pulses overlap.
This region is of particular interest as the resulting XUV pulses are in general shorter than the seed pulses\,\cite{finetti_pulse_2017} and hence would allow to study dynamics faster than the seed pulse duration. 
In the WPI study the XUV pulses were attenuated by a factor of $> 1000$ in order to avoid saturation of the transition by placing metal filters behind the amplification stage.
Instead one could reduce the gain in the amplification stage, the seed laser intensity and the dispersive strength,  while still keeping sufficient pulse energy for time-domain spectroscopy studies.

Note that the applicability of our work is not limited to the HGHG process.
Indeed, we expect a similar behavior in any harmonic generation process whenever saturation comes into play.
Hence another aspect of this work is the combination of the phase-modulation technique with table-top HHG sources. 
In fact, we have very recently combined the phase modulation technique with an HHG source\,\cite{wituschek_phase_2020}. In these  experiments we did not observe any saturation effects. The combination with HHG sources is particularly interesting since the achievable pulse durations can be much shorter than in HGHG, which in combination with the presented phase-modulation approach may pave the way for studies with sub-fs time resolution and high spectral resolution.
Furthermore, time-domain spectroscopy studies using notoriously weak XUV pulses generated in HHG setups would greatly benefit from the enhanced sensitivity of the phase-modulation method.

Yet another application of our work is the ability to characterize the underlying harmonic generation process using nonlinear ACs as a probe.
By comparison with simulations and by using well characterized seed pulses further information on the role of saturation, chirp and inhomogeneities in the generation medium may be inferred from the measured nonlinear interferograms.

\section*{Funding}
Funding by the Bundesministerium f\"ur Bildung und Forschung (BMBF) \textit{LoKo-FEL} (05K16VFB) and \textit{STAR} (05K19VF3), by the European Research Council (ERC) with the Advanced Grant \textit{COCONIS} (694965) and by the Deutsche Forschungsgemeinschaft (DFG) RTG 2079 is acknowledged. Also, funding by the Swedish Research Council and by the Knut and Alice Wallenberg Foundation, Sweden, is acknowledged

\section*{Acknowledgments}
We thank Maxim Gelin for providing the code used for the Runge-Kutta integration of the Schr\"odinger equation in Sec.\,\ref{sec:WPI}. We gratefully acknowledge the support of the FERMI staff.

\section*{Disclosures}
The authors declare no conflicts of interest.

\bibliography{literature/fermi_overlap_bib}

\end{document}